# Measurement of Neutrino's Magnetic Monopole Charge, Dark Energy and Cause of Quantum Mechanical Uncertainty

Eue-Jin Jeong,
The University of Texas at Austin
Austin TX USA

Dennis Edmondson
University of Washington Seattle
Seattle WA USA

## Abstract

Charge conservation in the theory of elementary particle physics is one of the best-established principles in physics. As such, if there are magnetic monopoles in the universe, magnetic charge will most likely be a conserved quantity like electric charges. If neutrinos are magnetic monopoles, as physicists have reported the possibility, the Earth should show signs of having magnetic monopole charge on a macroscopic scale since neutrons must also have magnetic monopole charge if general charge conservation principle is valid. To test this hypothesis, experiments were performed to detect the collective effect of magnetic monopole charge of neutrons on the earth's equator using two balanced high strength neodymium rod magnets. We were able to identify non-zero magnetic monopole charge of the individual neutrons from the experiments. The presence of individual magnetic monopole charges in the universe prompted proposition of the new symmetric form of Maxwell's equations. Based on the theoretical investigation of these new Maxwell's equations, we conclude that magnetic monopole neutrinos are the cause of the origin of quantum mechanical uncertainty, dark energy and the medium for electromagnetic wave propagation in space.

## Introduction

In 1931 P.A.M. Dirac theorized that for an electric charge to be quantized, a magnetic monopole must exist [1]. Years later, in 1974, Polyakov [2] and t' Hooft [3] discovered that the existence of monopoles follows from quite general ideas about the unification of the fundamental interactions.

Other pioneers in the field, Alan Chodos [4], E. Recami [5], O.M.P. Bilaniuk, V.K. Deshpande [6] and O.M.P. Bilaniuk, E.C.G. Sudarshan [7] had discussed the possible existence of tachyons. Some GUTs [8], such as Pati-Salam model [9] and superstring theory [10], predicted the existence of magnetic monopoles as well. E. Ricami [5] and J.J. Steyaert [11] discussed the possibility of neutrinos being tachyonic magnetic monopoles.

Recently the joint experimental team of UK, South Africa, Spain, France, and Brazil [12], published reporting that the lightest neutrinos have an upper bound mass of 0.086 $eV$ with a 95% confidence level. However, stationary neutrinos have not been detected either directly or indirectly, which is necessary to prove neutrinos are de facto ordinary matter particles. Also, there is no evidence of right handed neutrinos which are most likely to be observed if neutrinos can be stationary. Neutrinos have the unusual property of the mass oscillation between flavors which has not been observed in the electron families. Therefore the questions of whether neutrinos are tachyons, whether there are magnetic monopoles in the universe, and/or whether neutrinos are both tachyons and magnetic monopoles is still one of the great mysteries of the universe.

In conventional nuclear beta decay processes, a neutron decays into a proton, electron, and an anti-neutrino as follows.

$$_{0}^{1}n \rightarrow \,_{1}^{1}p + \,_{-1}^{0}e + \,_{0}^{0}\overline{\nu}_{e} \qquad (1)$$

The electric charge and baryon number are conserved in the process, while the neutrino carries the left-over energy and momentum. In this form of weak decay, the neutrino is an inactive subatomic particle because it does not have any type of charge and once it loses its energy there is no interaction of the neutrino with other particles, which is unusual from utilitarian philosophical perspective especially considering the massive number of their presence in the universe.

In this report, we present new experimental evidence that neutrinos are light magnetic monopoles by measuring the Earth's magnetic monopole charge originated from the massive number of neutrons based on the general charge conservation principle in particle physics.

## Experimental Principle

There are substantial numbers of neutrons in the atomic elements on Earth in their composite nuclear structures. In essence, the task of measuring the magnetic monopole charge of neutrino becomes the task of measuring the Earth's magnetic monopole charge of neutrons based on the assumption that the magnetic charge is conserved.

If a well-balanced long cylindrical high strength test magnet having a pivot at its midpoint is placed in the horizontal position near the equatorial surface of the earth, the assembly will tilt toward one side or the other depending on the polarity and the strength of the Earth's magnetic monopole charge.

The measurements was conducted by using two long permanent test bar magnets made of even number of cylindrical disc neodymium magnets having different, magnetic strength, diameters and lengths stacked into long a cylindrical form. The center of the bar magnet was placed on the pivot of a tightened string made of nonmagnetic material with negligible thickness to ensure minimum torque resistance on the test rod magnet to minimize obscuring the experimental data. Any minute tilting force on the sensor magnets is measured on the horizontal position using the precision micro scale to reflect the collective magnetic monopole charge of the neutrons on Earth.

The most critical task in the experiment is to separate the effect of Earth's dipole geomagnetic field on the balance of the detector magnet for the accurate measurement of the magnetic monopole charge of the Earth. Hence, it becomes a critical issue to choose the right location on the surface of the Earth to minimize the geomagnetic effect from getting into the measurement. As shown in Fig 1 and Fig 2, the location near the equator is necessary to avoid any vertical component of the dipole geomagnetic field getting into the data in the first order. Also, if the Earth had perfectly even density and uniform surface, any location near the equator would be acceptable in the second order. However the presence of deep sea ocean which is filled with water that has very low density of neutrons, high mountains that has large concentration of neutrons above the ground and volcanoes that has irregular density of underground matter can cause unnecessary errors in the measurement of the Earth's magnetic monopole field. Therefore, we decided to choose Cuenca as the location to perform the experiment since Quito is too close from the two volcanoes one in Pichincha 24 Km in the north and Cotopaxi 50 Km in the south. Cuenca is located 450 Km south of Cotopaxi and 104 Km east of the Pacific Ocean which is considered reasonably far outside of the geographical aberration.

The next critical factor is to balance the weight of the each halves of the bar magnet on both sides of the pivot because unless the test magnets are precisely balanced including the magnetic field strength, weights and length, measurement of the earth's magnet monopole strength would be inaccurate. The problem of balancing the weight of the bar magnet is that beside the actual weight due to the mass, there are magnetic forces affecting the measurement. This is achieved in the first order by choosing the identical number of the uniform sized disk magnets on both sides and second by selecting four disk magnets that have the identical magnetic strength measured by the precision magneto meter and place them at the end of each of the bar magnets since there are measurable minute differences in the field strength in some of the disk magnets.

Since the thickness and radius of each disk magnets are factory set, the accuracy of the length and weight of the both side of the bar magnets are extremely high with less than $\pm 1\%$ of error. The third step calibration is done by measuring the weight of each halves of the bar magnet by placing each separately in vertical position on the micro scale with N pole side down and by compensating any minute difference in weight by adding ten milligram range of weight in the middle of the lower weight side of the bar magnet. After the adjustment, both sides of the bar magnet turned out to have the identical mass, length and magnetic field strength within the error of $\pm 1\%$.

# Measurement and Analysis of Experimental Data

We chose the location (2.899350° S, 78.989264° W) in Cuenca Ecuador, a city close to the equator with high plateau, equipped with two sets of long neodymium cylindrical bar magnets, a precision magnetometer (WT10A), and a precision digital micro scale (AMOTGR 2001).

One of the assembled neodymium magnets has the length of 19 cm and the diameter of 10 mm, where both sides have magnetic fields of the same magnitude measured by the precision magnetometer, and the other neodymium dipole magnet has the length of 16.6 cm and the diameter of 12 mm. The earth as a monopole magnet is expected to exert magnetic force on these test magnets by pulling down on one side and pushing up the other when the balanced dipole magnet is placed horizontally resting on a tightened string of negligible thickness.

If consistent measurable tilting occurs on the two different precision balanced magnetic monopole sensors near the equator, it indicates that the earth has magnetic monopole charge stemming from the accumulative magnetic monopole effect of its neutrons.

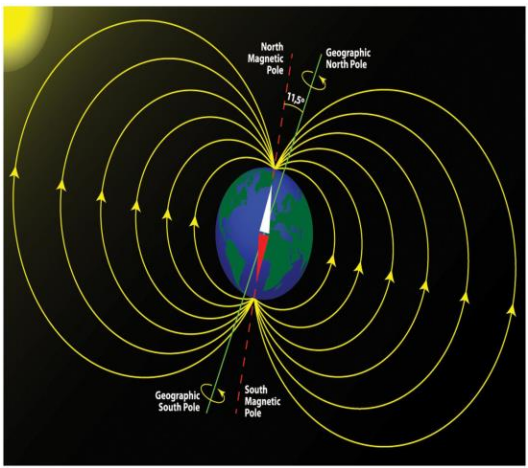
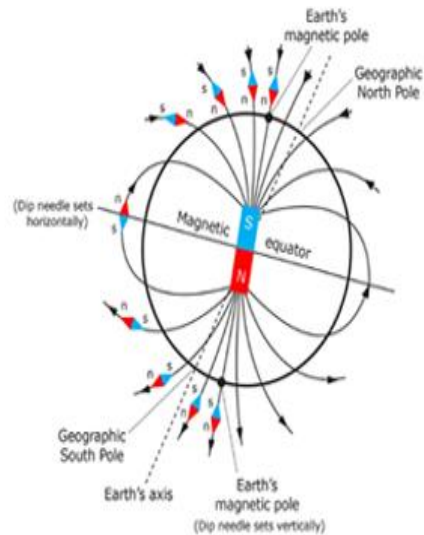

Fig 1 Curved Earth's geomagnetic field lines relative to the surface of the Earth [13]

Fig 2 Close up depiction of the Earth's geomagnetic field lines [14]

The close up depiction of the geomagnetic field lines in Fig 2 show that in the northern and southern hemispheres, there is strong vertical component of the geomagnetic field because the geomagnetic field lines converge toward the poles and it is no longer horizontal to the surface of the Earth except near the equator. This vertical component of the geomagnetic field can not be distinguished from the central monopole magnetic field originated from the neutrons of the Earth in the measurement and this is the reason we chose the equatorial region of the earth to perform the experiment.

The measurement of the magnetic field on the ground of the earth also has the specific advantage that the data are not affected by the parasitic magnetic field in the ionosphere caused by the solar wind and ionized particles along the path of the satellite.

The basic equation of the force between two different permanent magnet poles $q_{m1}$ and $q_{m2}$ is given by

$$F = \frac{\mu_0 q_{m1} q_{m2}}{4\pi r^2} \quad (2)$$

, which is formally equivalent to the equation for the force between two different electrostatic charges [15].

To determine the magnetic strength of the test magnets, $F_1$ is measured at the moment when each halves of the full length of the test magnets are pulled to be separated at the preset separation gap distance $r_1$. From the randomly chosen separation distances of $r_1$ 5.2 mm and 5.1 mm, 1.2 kg and 0.65 Kg of weight equivalent horizontal magnetic pulling forces were measured respectively (5)

$$F_1 = \frac{\mu_0 q_m q_m}{4\pi r_1^2}. \quad (3)$$

to estimate the magnetic charge strength of the two test magnets. It was determined that using this method to measure the strength of each magnetic pole is more direct and reliable than using the result of magnetic flux density measured by Gauss meter and convert it into the magnetic charge strength using the mathematical relation. This method avoids the calibration dependency error caused by the magnetometers that use the strength of the earth's geomagnetic field that changes depending on time and location for calibration.

The next step is to measure the strength the magnetic monopole charge of the earth and calculate the estimated number of neutrons in the entire earth to obtain the single magnetic monopole charge of the individual neutron and subsequently that of the individual neutrinos.

For the total magnetic monopole charge $Q_m$ of the Earth, the tilting force due to the interaction between the Earth magnetic monopole and the dipole magnet placed on the horizontal pivot at the center is given by

$$F_2 = \frac{2\mu_0 q_m Q_m}{4\pi R^2} \quad (4)$$

The factor of 2 comes from the two sides, one from the attractive force between N-S and the other from the repulsive force between N-N on the opposite side of the test magnet. The elevation at the test site in the city of Cuenca Ecuador is 2.56 km above sea level.

Therefore, $R = (6368 + 2.56) Km$ and the downward tilting weights measured repeatedly at the horizontal position of the magnets turned out to be $(0.78 \pm 0.04)g$ and $(0.52 \pm 0.03)g$, respectively, on the digital micro scale for the two test magnets.

**Measurement Data within the Error of $\pm 5\%$**

| Rod Magnet | length | diameter | $F_1$ | $r_1$ | $F_2$ | $B_0$ | Weight R/L |
|---|---|---|---|---|---|---|---|
| Rod Magnet 1 | 16.6 cm | 12 mm | 1.2 kg | 5.2 mm | 0.78 g | 454 mT | 71.83 g/71.84 g |
| Rod Magnet 2 | 19 cm | 10 mm | 0.65 kg | 5.1 mm | 0.52 g | 413 mT | 56.83 g/56.85 g |

(5)

The south pole side of the magnet tilts downward consistently for both test Rod Magnet1 and Rod Magnet2, indicating that the Earth is a north magnetic monopole. "Weight R/L" is the weight of half of the each test magnet separately measured before assembling the two halves into the full length.

**Analysis**

Using the data collected after repeated measurements, we found $Q_m = (2.75 \pm 0.14) \times 10^{16} Weber$ for the Earth's monopole magnetic charge of the north kind measured by Rod Magnet1 and $Q_m = (2.54 \pm 0.13) \times 10^{16} Weber$ for the same north kind measured by Rod Magnet2. The average of the two measured values $Q_m = (2.65 \pm 0.13) \times 10^{16} Weber$ is determined for the north magnetic monopole charge of the Earth.

The reason for employing the two different sets of test magnets was to eliminate spurious experimental results when only a single set of test magnets is used. The two permanent magnet bars are not of the same brand or of the same diameter or the length. Despite the deliberate choice of the two entirely different sets of test magnets, the experimental results were unexpectedly close from each other.

To estimate the magnetic field at the surface of the earth stemming from the measured magnetic charge of the earth $Q_m = (2.65 \pm 0.13) \times 10^{16} Weber$ at the distance $R = (6368 + 2.56) Km$ we use the magneto static force equation (2)

$$F = \frac{\mu_0 Q_m q_m}{4\pi r^2} \quad (6)$$

where $Q_m$ represents the total magnetic charge of the earth and $q_m$ a test magnetic monopole charge placed on the surface of the earth, and since $\mu_0/4\pi = 10^{-7}$, the magneto-static force is given by

$$F = 10^{-7} \frac{2.65 \times 10^{16} \times q_m}{(6370.56 \times 10^3)^2} = 65.3 \mu T \times q_m \tag{7}$$

This result shows that the strength of the magnetic monopole field at the surface of the earth at the equator due to the magnetic monopole charge of the earth is estimated to be around $65.3 \mu T$. The past measurement [16] of the earth's geomagnetic field from SWARM satellite was in the range between $25 \mu T$ near the equator and $65 \mu T$ near the pole. The satellites were actually measuring both the Earth's magnetic monopole and the geomagnetic dipole field together at the orbit since there are no means to measure them separately especially at the poles The Earth's magnetic monopole field strength adjusted at the height of 510 Km above the sea level is about $56 \mu T$ according to the result (7). This is very close to the generally accepted Earth magnetic field strength $50 \mu T$ measured from the satellites. Since the two methods of the earth magnetic field measurement are completely independent from each other, the agreement in the order of the magnitude of the earth magnetic field strength at the equator is considered significant. It is noted that we used the magnetometer only for the purpose of choosing the identical disk magnet to be placed at the end of the test bar magnet to make sure the magnetic fields at the tips are identical in strength. The minor difference could be caused by the fact that the laboratory magnetometers used in the satellite are calibrated based on the geomagnetic field of the earth which is known to fluctuate depending on time and location.

To estimate the total number of neutrons on Earth to calculate the individual magnetic monopole charge of the neutron, the element table [17] was utilized.

According to the element table, the following distribution applies for 99.9 percent of the mass on Earth's elements.

| %      | Element   | # Neutron/Proton | Weighted Average N/P |
|--------|-----------|------------------|----------------------|
| 5.63   | Iron      | 30/26            | 168.9/146.38         |
| 46.1   | Oxygen    | 8/8              | 368.8/368.8          |
| 28.2   | Silicon   | 14/14            | 394.8/394.8          |
| 2.33   | Magnesium | 12/12            | 27.96/27.96          |
| 8.23   | Aluminum  | 14/13            | 115.22/106.99        |
| 4.15   | Calcium   | 20/20            | 83/83                |
| 2.36   | Sodium    | 12/11            | 28.32/25.96          |
| 2.09   | Potassium | 20/19            | 41.8/39.71           |
| 0.565  | Titanium  | 26/22            | 14.69/12.43          |
| 0.095  | Manganese | 30/25            | 2.85/2.375           |
| 0.14   | Hydrogen  | 0/1              | 0/0.14           (8) |

The mass difference due to the isotopes is included in each element's average atomic mass. Subsequently we find the percentage distribution of the mass of the Earth to be 50.7% neutron and 49.3% proton, and the contribution from electrons is negligible. Using the known total mass of the Earth $5.972 \times 10^{24} Kg$ and the mass of a single neutron $1.675 \times 10^{-27} Kg$, the best estimated number of neutrons on the Earth turned out to be $\#n_{earth} = 1.8076 \times 10^{51}$. From these results, the single magnetic monopole charge $m_v$ of a neutron

$$m_v = \frac{Q_m}{\#n_{earth}} = (1.46 \pm 0.07) \times 10^{-35} Weber. \qquad (9)$$

is determined with 95% of confidence level. This level of the strength of the magnetic charge of the neutron is out of the range by the single neutron magnetic monopole charge detection method that may have been tried in the laboratory. Neutrons pass through a single slit and subjected to a strong magnetic field perpendicular to their path may be used to detect the deviation of their path from the one without the magnetic field. However, there is no report of this type of experiments were performed on neutrons. The key issue in such experiment would be in the possibility of obtaining the highest strength of the perpendicular magnetic field vs. the accuracy of the measurement of the deviation of the neutron's path from the one without the magnetic field.

Experiments were also performed at the opposite side of the globe in Pontianak (0.03109° S, 109.32199° E) Indonesia, Bangkok (13.7524° N, 100.5507° E) Thailand and Chengdu (30.6593° N, 104.0598° E) China. The same north magnetic monopole charge of the Earth was observed in Pontianak 0.03109° S. The experiment performed in Bangkok showed zero tilting effect on the balanced test bar magnets indicating that 13.7524° N latitude is where the effect from the north magnetic monopole effect of the earth cancels the south magnetic pole component of the Earth's geomagnetic field. In Chengdu 30.6593° N, strong geomagnetic field effect of the south-pole component (note: Earth's North Pole is magnetic south-pole) was observed overriding the monopole effect.

Further investigation showed that the repulsive monopole magnetic force between two matter objects is $1.14 \times 10^{-13}$ times smaller than the attractive gravity force between them, assuming that the same number of neutrons comprised each of the two matter objects. This indicates the repulsive magnetic monopole forces among the stars and planets are extremely weak and negligible compared to that of gravity on a planetary or a galactic scale.

The noted discrepancy of this finding from the results of past studies is that the measured magnetic monopole charge of the neutron does not match Dirac's prediction of $g = \frac{N}{2}\frac{\hbar c}{e}$ [18], where $h$ is Planck's constant and $N$ is an integer resulting in the calculated Dirac's magnetic monopole charge $N \times 0.99 \times 10^{-7}$.

However, if Dirac's magnetic monopole is taken for the magnetic monopole charge of a neutron, the repulsive magnetic force between matter objects becomes too large to be ignored in celestial mechanics. Newtonian mechanical description of planetary motion does not work in such cases.

There is a difference in the order of $10^{28}$ between Dirac's and the present report of the magnetic monopole charge $1.46 \times 10^{-35} Weber$. Dirac's magnetic monopole charge is close to the value $g = N \frac{\mu_0}{4\pi}$, where $\mu_0 = 4\pi \times 10^{-7} H/m$ is vacuum permeability.

It turned out that the measured magnetic monopole charge of a neutron coincides with the value of vacuum permittivity divided by Avogadro's number

$$8.854 \times 10^{-12} (Faraday/Meter)/6.022 \times 10^{23} = 1.47 \times 10^{-35} Weber \tag{10}$$

The possible reason is because while the vacuum permittivity $\varepsilon_0$ is generally perceived to be associated with electric field, the Coulomb electric field is actually inversely proportional to $\varepsilon_0$ because of the $\frac{1}{4\pi\varepsilon_0}$ factor while $\varepsilon_0$ and $\mu_0$ are mutually constrained by the relation $c = \frac{1}{\sqrt{\varepsilon_0 \mu_0}}$, which suggests that although $\varepsilon_0$ has the unit $Faraday/Meter$, the vacuum permittivity is directly related to the strength of the magnetic monopole charge density in space.

## Neutron Beta Decay Process Including Magnetic Monopoles

Based on the principle of the charge conservation in the universe including both the electric and magnetic charges, the full neutron beta decay process can now be presented by

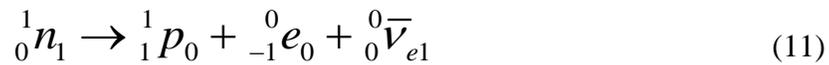

$$^1_0n_1 \rightarrow ^1_1p_0 + ^0_{-1}e_0 + ^0_0\overline{v}_{e1} \tag{11}$$

where the low right-hand side sub-indices indicate the number of conserved magnetic monopole charges. In this picture, W boson is identified as a temporary composite transient particle that has both the magnetic and electric charges before splitting into the electron and antineutrino. As J.J. Steyaert pointed out in his paper [4], the weak interaction in the standard model could be a manifestation of the magnetic monopole effect in the nuclear interaction processes. This new result of the existence of nonzero magnetic monopole charge prompts Maxwell's equations need to be modified to include the full spectrum of magnetic monopole phenomena that manifest in the universe.

## Symmetric Form of Maxwell's Equation

We propose the symmetric form of Maxwell's equations (12) which is based on the experimental confirmation of the magnetic monopole in the form of neutron's static charge and the magnetic current of the traveling neutrinos,

$$\nabla \cdot \vec{E} = \frac{\rho_e}{\varepsilon_0}$$

$$\nabla \cdot \vec{B} = \mu_0 \rho_m$$

$$\nabla \times \vec{E} = -\frac{1}{\varepsilon_0}\frac{\partial \vec{B}}{\partial t} - \frac{1}{\varepsilon_0}\vec{J}_m \quad (12)$$

$$\nabla \times \vec{B} = \mu_0 \vec{J}_e + \mu_0 \frac{\partial \vec{E}}{\partial t}$$

, where $\rho_m$ is the static magnetic monopole charge density formed by neutrons and $\vec{J}_m$ is the magnetic monopole current density from traveling neutrinos. Other researchers have reported similar forms of symmetric Maxwell's equation in the past [19]. The conventional expression $\nabla \cdot \vec{B} = 0$ in the original Maxwell's equation passed experimental test without measurable inconsistency because magnetic charge density $\rho_m$ is extremely weak and individually undetectable unless collective effect from massive number of neutrons is subjected to the test. The current density $\vec{J}_m$ in the third equation does not have a net value in free space because of the isotropy of the neutrino's flux, despite massive number of their presence in the universe which explains the absence of this term in the original Maxwell's equation.

Magnetic current is observable in the form of magnetic flux when coherent collimated stream of magnetic monopole neutrinos passes along the magnet, forming an air gap loop. The actual effect of the electrical current in a typical copper magnet wire according to equation (12) is to rearrange the flow of the random magnetic current that already exists in space into a coherent collimated stream instead of creating the entire magnetic flux from the empty space.

Maxwell's equations have not been found to violate any known physical phenomena while questions still remained on why the equations are not symmetric with respect to the magnetism and why there are no individual magnetic monopoles.

## Atomic Stability Condition and the Speed of Magnetic Monopole Neutrinos in the Universe

The symmetric Maxwell's equation (12) provides a new solution for the electric field created by a traveling magnetic monopole neutrino which is given by

$$\vec{E} = -\frac{1}{4\pi\varepsilon_0} \frac{m_v \vec{\upsilon} \times \hat{r}}{r^2} \tag{13}$$

where $m_v$ is the single magnetic monopole charge and $\vec{\upsilon}$ is its velocity. This result is a mirror image of the magnetic field created by moving electric charge in the original Maxwell's equations. According to the classical electro-dynamic description of the atoms, the orbiting electrons radiate photons, lose energy and collapse into the nucleus, which was the cause of the development of quantum mechanics. The core of the puzzle was how the electrons stay afloat from the nucleus and avoid collapse especially the ones in s orbital that has zero angular momentum assuming that the exchange of photon energy allows the electrons with nonzero angular momentum to stay in the orbit. The quantized photonic energy relation $E = h\nu$ was originated from the property of the vacuum space without reference to the presence of electrons nearby. As such, there is a possibility that some unknown external electric fields in space could counteract the attractive Coulomb force between the nucleus and the electron. In fact, the electric field (13) created by traveling magnetic monopole neutrinos could be a candidate for providing the source of such external electric field.

In the absence of a viable mechanical picture of the electron's behavior in atoms, we propose that the electric fields (13) created by traveling magnetic monopole neutrinos in space are responsible for the stability of the atomic structures. The validity of this proposition depends on its ability to address the physical details in the understanding of the universe in accordance with the numerical and observational data.

To maintain the atomic stability, the vacuum electric field (13) created by the traveling magnetic monopole neutrinos must have equivalent strength as the Coulomb electrostatic field of the proton

$$E = \frac{1}{4\pi\varepsilon_0} \frac{e}{r^2} \tag{14}$$

to prevent the electron from collapsing into the proton based on the electro-dynamic principle.

The two electric fields (13) and (14) are of different geometrical types since one (13) is tangential in cylindrical geometry and the other (14) is radial in spherical form. However, depending on the frequency of large number of directionally varying magnetic monopoles passing by the proton, the two fields counteract each other and keep the electron in a constant agitating and rotating state. The unique combination of these two geometrically different electric fields creates enormous complexity due to the massive number of randomly traveling background neutrinos. In all these cases, unless the order of magnitude of the strengths of the electric field (13) and (14) are within the same range, the electron will either collapse into the proton or fly away out of the boundary of the atomic orbit.

By equating the electric field (13) at $\theta = 90°$ ($\vec{\upsilon} \perp \hat{r}$) and (14), the optimum speed of the magnetic monopoles to prevent the collapse of the electron-proton substructure turns out

$$\upsilon = \frac{e}{m_v} \quad (15)$$

For the charge of an electron $e = 1.6021766 \times 10^{-19} Coulomb$ and the measured magnetic monopole charge $m_v = 1.46 \times 10^{-35} Weber$, we find the speed of the magnetic monopole neutrino $1.098 \times 10^{16} m/\sec$, which is $3.655 \times 10^7$ times the speed of light to be able to stabilize the hydrogen atom and/or all the atoms since there is no dependency on $r$ in the relation (15).

This result indicates that the calculated speed of the background neutrinos that is needed to stabilize the atomic structure is universally identical regardless of the specific atomic numbers or where the atom is located, and this particular speed of neutrino's travel is necessary to maintain the stability of the entire architecture of the material universe without having to resort to quantum mechanical uncertainty.

In fact, the negative mass squared problem of the neutrino experiments reported by the five major institutions [18] has prompted the suspicion that, after all, neutrinos may not be ordinary matter particles. Stationary neutrinos have not been detected so far which is necessary to prove neutrinos are ordinary matter particles. Also, there is no evidence of right handed neutrinos which are most likely to be observed if neutrinos can be stationary. Neutrinos have the unusual property of the mass oscillation between flavors which has not been observed in the electron families. Despite the recent success of the detection of the nonzero rest mass of neutrinos [12], there are too many inconsistencies to fit neutrinos into the category of the ordinary inside light cone particles and that the simplest possible way to resolve the mystery of the neutrinos may be to change our point of view and determine that neutrinos are actually tachyons. The most glaring inconsistency is in the fact that the vast majority of the $10^{79}$ neutrinos [20] in the universe are not detectable and few of the detected neutrinos have their speed close to the speed of light which doesn't necessarily prove that the rest of the neutrinos would also be traveling close to the speed of light since neutrinos, as fermions, are not required to have the fixed speed of travel. If most of them are not stationary, based on the fact that stationary state neutrinos have not been detected, the only possibility is that the vast majority of them must be traveling extremely faster than the speed of light which is consistent with the negative mass squared problems reported by the earlier experimental results. As such, despite the questionable possibility that the background neutrinos could travel $3.655 \times 10^7$ times the speed of light, we proceed to investigate further to see what else can be found from the same proposition.

## Dark Energy, Vacuum Electric Field and Quantum Mechanical Uncertainty

The vacuum electric field $\vec{E}(\vec{r},t)$ created by the fast-traveling magnetic monopole tachyonic neutrinos in the background is given by

$$\vec{E}(\vec{r},t) = \sum_{i=1}^{N} \frac{1}{4\pi\varepsilon_0} \frac{m_v \vec{v}_i \times (\vec{r} - \vec{r}_i(t))}{|(\vec{r} - \vec{r}_i(t))|^3} \quad (16)$$

which is a multiple sum of the individual electric fields (13), where $m_v$ is the magnetic monopole charge of the neutrino, $\vec{v}_i$ is its velocity, $\vec{r}_i(t)$ is the position of the particle at time $t$ and $N$ is the total number of neutrinos in the universe which is estimated to be in the order of $10^{79}$ [21].

The uniformly distributed superluminal magnetic monopole neutrinos accumulate repulsive magnetic potential energy in space which is written in differential form

$$dU_r = \frac{\mu_0 \rho_m \left(\frac{4}{3}\pi r^3\right)}{4\pi r} d\left(\rho_m \left(\frac{4}{3}\pi r^3\right)\right) \quad (17)$$

, where $\rho_m$ is the magnetic monopole charge density in the universe.

Calculation of $U_r$ over the radius of the observable universe identified the repulsive magnetic potential energy from the uniformly distributed magnetic monopole neutrinos given by

$$U_r = \frac{3\mu_0 N^2 m_v^2}{20\pi R} \, Joule \quad (18)$$

where $R$ is the radius of the observable universe $4.4 \times 10^{26}\, meter$ [22], $N$ is the total number of magnetic monopole neutrinos which is in the order of $10^{79}$ and $m_v$ is the single magnetic monopole charge $1.46 \times 10^{-35}\, Weber$ and $\mu_0 = 4\pi \times 10^{-7}\, H/m$, and the vacuum energy in this particular case is $U_r = 2.907 \times 10^{54}\, Joule$.

Since nature tends to move toward the lowest energy state whenever possible, the universe has to expand to reduce its repulsive vacuum potential energy according to the result (18). There is also energy created by the interaction of the electric fields carried along by the traveling magnetic monopoles represented by the expression (13). For example, when two neutrinos travel in parallel trajectory, there is repulsive electric force perpendicular to both the direction of their travel and the line connecting the two neutrinos because the directions of the two cylindrical electric fields run against each other. Also, when two neutrinos travel in opposite direction passing each other in parallel trajectory, there is attractive interaction perpendicular to both the direction of their travel and the line connecting the two. As such, any type of interactions between the electric fields generated by traveling magnetic monopoles does not directly contribute to either attraction or repulsion among the neutrinos themselves while they still create energy. In general, energy is related to either attraction or repulsion among the material objects in both the case of gravitation and static electromagnetism but this is a new form of energy because the lines of force do not connect the two objects in action.

The energy created by these interactions is represented by

$$dU = \frac{m_\nu n \upsilon \frac{4}{3}\pi r^3}{4\pi\varepsilon_0 r} d\left(m_\nu n \upsilon \frac{4}{3}\pi r^3\right) \qquad (19)$$

where $n$ is the number density of neutrinos in space, $m_\nu$ is neutrino's magnetic monopole charge and $\upsilon$ is the speed of the traveling background neutrinos. By integrating (19) over the radius $R$ ($4.4\times10^{26}\,meter$) of the observable universe, the total electric vacuum energy of the universe is found

$$U_{electric} = 3.13\times10^{103}\,Joules \qquad (20)$$

We identify this energy as the main part of the dark energy that does not directly contribute to the expansion but there to support the structural integrity of the material universe. From the perspective of quantum field theory, the fact that the force lines of this interaction do not connect directly the two objects in motion yet they accumulate large amount of energy in space is consistent with the fact that photons as bosons in quantum field theory do not obey Pauli's exclusion principle since they are created by the traveling tachyonic magnetic monopoles (fermions) as spiraling electric fields according to Maxwell's equation.

Using the relation $\dfrac{U}{\frac{4}{3}\pi R^3} = \dfrac{1}{2}\varepsilon_0 E^2$ we obtain the strength of the vacuum electric field (16) given by

$$|E| = 1.41\times10^{17}\,Newton/Coulomb \qquad (21)$$

Considering the rigidity of the medium determines the upper limit of the frequency of the waves it carries in general, this is considered a strong vacuum electric field as a medium to handle ultra high frequency electromagnetic waves propagating in space. The rapidly fluctuating background electric field has the average strength same as the Coulomb electric field at the distance $10^{-13}\,m$ from the proton which is larger than proton's charge radius $8.4\times10^{-16}\,m$ but smaller than Bohr radius $5.3\times10^{-11}\,m$, which serves both as a mechanism for quantum fluctuation and also as a medium for electromagnetic wave propagation in space.

The equation of motion of a free electron in vacuum using the vacuum electric field (16) is written,

$$m_e \frac{d^2\vec{r}}{dt^2} + \frac{e}{4\pi\varepsilon_0} \sum_{i=1}^{N} \frac{m_v \vec{\upsilon}_i \times (\vec{r} - \vec{r}_i(t))}{|(\vec{r} - \vec{r}_i(t))|^3} = 0 \qquad (22)$$

, which defines the inertial mass of the electron in the vacuum. The motion of a charged particle in free space is restricted by the presence of the rigid fluctuating electric field, yet the particle's position and momentum are not precisely determined on the microscopic scale.

The quantum particle's position and momentum in free space are under constant interference of the magnetic monopole current of the neutrinos that have random characteristics in motion. This indicates that in addition to the contribution to dark energy, the stochastic interaction between the traveling magnetic monopole neutrinos and charged particles causes the quantum mechanical uncertainty of position and momentum without apparent external forces of known origin.

Subsequently we write the full equation of motion of an electron in an isolated hydrogen atom in space as

$$m_e \frac{d^2\vec{r}}{dt^2} + \frac{e}{4\pi\varepsilon_0} \sum_{i=1}^{N} \frac{m_v \vec{\upsilon}_i \times (\vec{r} - \vec{r}_i(t))}{|(\vec{r} - \vec{r}_i(t))|^3} + \frac{e^2}{4\pi\varepsilon_0 r^2} = 0 \qquad (23)$$

where $m_e$ is the mass of the electron and $e$ the single electronic charge.

Equations (22) and (23) must be reduced into probabilistic statistical forms to obtain meaningful physical information on the electron because there are no means to predict the precise motions of each and every individual tachyonic magnetic monopole neutrinos in the universe other than the physical constraints that they are expected to observe collectively. A mathematical reduction of the equations (22) and (23) into probabilistic forms, the task of which is out of the scope of the present report, is expected to result into a formal structure reminiscent to Schrödinger equation when the following constraints are applied.

In general, fermions do not have a fixed speed of travel and a case was projected where tachyonic magnetic monopole neutrinos have a Maxwell-Boltzmann type velocity distribution like gas molecules peaking around $1.095 \times 10^{16} m/\sec$ and tapering off to zero at both the speed of infinity and speed of light at the low end. However, since tachyonic neutrinos do not have inertial mass, such as matter particles, nor their motion is influenced by ambient temperature, it is concluded that their velocity distribution must be radically different and close to a delta-function peaking at $1.095 \times 10^{16} m/\sec$ in addition to the isotropy condition $\sum_{i=1}^{N} m_v \vec{\upsilon}_i = 0$.

On the other hand, the arguments against the legitimacy of quantum mechanics, for example, action at far great distance, non-locality, quantum entanglement, hidden variables and incompleteness of quantum mechanics [23], are expected consequences of the theoretical efforts to incorporate the inherent interference from the superluminal tachyonic magnetic monopoles in space that is missing in the local probabilistic description of the classical quantum mechanics since $3.655 \times 10^7$ times the speed of light is almost instant to reach the far side of the universe in practical sense.

## Discussion

Prediction of the expanding universe and the calculation of the dark energy using the quantum stability condition (15) were not the intended objective at the start of this investigation. We were investigating the mechanical stability aspects of the atoms using the solution from the new symmetric Maxell's equations based on the measured magnetic monopole charge of the neutrinos. The behavior of the electrons in s-orbital prompted the proposition of the atomic stability condition using the new solution of the electric field created by traveling magnetic monopoles.

The incredibly fast speed of travel of these particles assures their uniform density in the vast space and it creates large sums of vacuum energy due to the interactions among the neutrino's velocity induced electric fields (13). The existence of the stable atomic structures and the material universe is not compatible without such a paradoxical arrangement since the undetermined mechanical aspect of the quantum uncertainty is supported by the electric fields created by the traveling background magnetic monopole neutrinos, which is the basis of the existence of the material universe.

In a most recent development, neutrinos were observed to change flavors while traveling in space, especially when they emerged out of the South Pole of the Earth according to the recent experimental report [24]. Since the neutrino's magnetic monopole is north kind, the background neutrinos in space are pulled into Earth's North Pole (which is magnetic south-pole), pass through the Earth's rotation axis and exit South Pole with changed flavor and gained energy, which explains the experimental result. This means the neutrinos that disappeared into the background after losing energy came back to be detected due to the energy-exchange interactions between the traveling north magnetic monopole neutrinos and the Earth's geomagnetic field. This experimental result suggests that it is not the mass but the magnetic monopole charge of the neutrinos that activates neutrinos to gain energy and change flavors. It was also reported that the solar wind tends to move more toward North Pole than toward South Pole which was a mystery. The mystery may be resolved if the two neutrons (as north magnetic monopoles) inside the helium ion nuclei which comprises one of the major components of the solar wind are attracted toward North Pole (south magnetic pole) and repelled by South Pole (north magnetic pole).

# Conclusion

We tested the hypothesis that neutrons could have magnetic monopole charge by measuring the collective effect from the massive number of neutrons inside the earth on the two accurately balanced test bar magnets. Following the definite non-zero result of neutron's magnetic monopole charge from the measurement, we generalized Maxwell's equation to symmetrical form since the charge conservation principle dictates neutrinos also must have magnetic monopole charge.

Assuming that quantum particles must have a certain level of kinematical aspects in their behavior in addition to the statistical one, we postulated that many of the background neutrinos should affect the dynamics of the electrons orbiting the nucleus. This was considered natural since traveling magnetic monopoles generate spiraling electric field on their paths according to the symmetric form of Maxwell's equation. The strength of the electric field created by the traveling magnetic monopoles should be comparable to the binding energy of the electron in atomic structures in s-orbit to guarantee a stable atomic structure and consequently the universe. From this result, we concluded that the stochastic nature of the quantum phenomena is due to the random motion of the background neutrinos creating chaotic electric field in vacuum, "stabilized" due to their uniform density. This pervading invisible and undetectable random electric field in space is also identified as the essence of the medium for the electromagnetic wave propagation known as Aether. We subsequently obtained the mathematical expression for the energy created by the pervading electric field caused by traveling neutrinos which matches with the estimated dark energy reported by other researchers.

The speed of neutrinos was calculated without the prior restriction from the result of special relativity because Maxwell's equation has no previous record of having failed the experimental test. While the speed of light limit applies to matter particles that obey Newtonian mechanical principle, there is no evidence showing that neutrinos have the same mechanical property as matter particles that follow Newton's law of motion since stationary neutrinos have never been found despite the sheer number of their existence in the universe. The previous measurement of the non-zero mass of the neutrinos by experiment doesn't necessarily prove neutrinos can be stationary. Since the first law of Newtonian mechanics states that "a body continues in its state of rest, or in uniform motion in a straight line, unless acted upon by a force", there must be verifiable existence of stationary state, for the object can be subjected to Newtonian mechanical principle. Absence of stationary state of neutrinos means neutrinos can not be subjected to special relativity and therefore the imposition of the speed of light limitation onto neutrinos, after all, was an overreach generalization that it didn't have its original compass. Also, from the practical technological point of view, the property of the vacuum space filled with tachyonic magnetic monopole neutrinos could provide us with a clue on how to maneuver the vast space of the universe without having to rely on the conventional propulsion technology in the future considering that the understanding of the physical property of the medium that needs to be passed is the first step to find the key for the necessary technology.

## Acknowledgement

It was after accidentally noticing the significant tilting of the compass needle in the high latitude geographical regions of the Earth in 2017 that we suspected there may be some kind of geocentric magnetic field that affects the balance of the compass needle since the compass manufacturers will not roll out their compasses in such an unbalanced state. Since we have wondered why there are no magnetic monopoles in the universe, it came to the conclusion that, if there is magnetic monopole field emanating from the Earth, the tilting effect of the balanced test bar magnet due to the vertical component of the geomagnetic field must be isolated by all means to accurately account for the Earth's collective magnetic monopole field originating from all the neutrons. This was the course of event that spurred the decision for the experiment to detect the earth's magnetic monopole field. The first author is grateful to his mentors who taught him the wonders of mathematics that predicts the exact nature of the physical phenomena.